\documentclass[]{qjmam}

\usepackage{amssymb}
\usepackage{amsmath}
\usepackage{caption}
\usepackage{graphicx}
\usepackage{color}

\startpage{1}
\yr{}
\vol{}
\issue{}

\DeclareMathAlphabet\mathbit
    \encodingdefault\rmdefault\bfdefault\itdefault
\DeclareOldFontCommand{\bi}{\normalfont\bfseries\itshape}{\mathbit}

\def\dd{\mbox{d}}
\def\bea{\begin{eqnarray}}
\def\eea{\end{eqnarray}}

\newcommand{\pr}{\sigma}
\newcommand{\gr}{\mathrm{Gr}}

\newcommand{\no}{\noindent}
\newcommand{\be}{\begin{equation}}
\newcommand{\ee}{\end{equation}}

\def\fakebold#1{\relax\ifvmode\leavevmode\fi%
\ifmmode%
\setbox0=\hbox{$#1$}%
\else%
\setbox0=\hbox{#1}%
\fi%
\kern-.02em\copy0 \kern-\wd0%
\kern .04em\copy0 \kern-\wd0%
\kern-.0125em\raise.02em\box0%
}%

\def\dd{\mbox{d}}
\def\pl{\partial}
\def\bea{\begin{eqnarray}}
\def\eea{\end{eqnarray}}

\def\edt{\mbox{e.d.t.}}

\usepackage[numbers,round]{natbib}

\makeatletter
\renewcommand\@biblabel[1]{{\bf #1}.}
\renewcommand\@cite[1]{({\bf #1})}
\makeatother

\begin{document}

\title[{mixed~convection~on~a~rotating~disk}] 
{MIXED CONVECTION ABOVE A ROTATING DISK}

\author[J. Arrieta-Sanagust\'{i}n, M. G. Blyth] {JORGE ARRIETA-SANAGUST\'{I}N \footnote{Permanent address: Departamento de Ingenier\'{i}a T\'ermica y de F"uidos, Universidad Carlos III de Madrid, Legan\'{e}s, Spain. } and M. G. BLYTH}



\address{School of Mathematics, University of East Anglia, Norwich, UK}

\received{\recd . \revd }

\maketitle

\eqnobysec

\begin{abstract}
Mixed convection above a horizontal disk rotating in a semi-infinite fluid is examined when the disk is heated
so that its temperature varies quadratically with distance away from its centre. Steady similarity solutions
are presented for a range of values of the two dimensionless parameters, a Prandtl number and a Grashof number.
The results are corroborated by asymptotic analyses undertaken in the limits of small and large Prandtl number. 
For finite Prandtl number, the existence of multiple solutions at fixed Grashof number is revealed. 
The similarity structure for steady solutions fails beyond a critical Grashof number and this is interpreted 
in terms of a finite-time singularity of the unsteady form of the governing equations.
\end{abstract}


\section{Introduction}

Free convection refers to the motion created by thermal effects in a fluid which is otherwise at rest. 
Forced convection refers to the situation in which an 
external flow field, such as a uniform stream, directly controls the movement and distribution of heat
within a fluid. Of particular interest in the present work is the case of 
mixed convection in which free-convective effects interact directly with a forced convective motion.

In a typical free convection problem, buoyancy forces drive motion directly. This occurs, for example, when 
a heated plate is held inside a fluid in a vertical position. However, fluid motion may result
from thermal buoyancy effects even when no component of the buoyancy force acts in the direction of the 
subsequent motion.
Stewartson \cite{STEW58} and Gill {\it et al.} \cite{GILL65} 
demonstrated that free horizontal convection may occur in a boundary layer on a semi-infinite horizontal plate at a 
constant temperature. In this case there is no component of the buoyancy force
in the direction of motion; rather the horizontal flow occurs indirectly as a result of a pressure gradient induced along the 
plate. Amin \& Riley \cite{Amin1990} demonstrated that horizontal free convection takes place above an infinite plate 
held in a quiescent fluid when the plate temperature 
varies along its length. This flow is also of boundary-layer type provided that the 
Grashof number, which measures the relative strength of the buoyancy-induced convection to viscous effects, is large. 

In a study of mixed convection at a vertical plate held at a constant temperature, 
Merkin \cite{MERK} showed that when the buoyancy force
opposes the prevailing direction of motion,  the boundary layer eventually separates and the 
solution fails at a singularity (akin to the classical Goldstein singularity in an isothermal boundary layer) 
where the skin friction vanishes. Similar singular behaviour and flow separation is seen 
for uniform flow past a flat plate with constant heat flux \cite{WILKS, WILKS2}, and a discussion of the character
of the singularity at the separation point was provided by Hunt \& Wilks \cite{HW80}.
Jones \cite{JONES} showed that for a heated semi-infinite plate inclined at a small negative angle to the 
horizontal, flow separation occurs without meeting a singularity, and it is possible to continue the integration of 
the boundary-layer equations downstream. The absence of a singularity may be attributed to the ability of the
pressure-gradient to adjust interactively to accommodate the flow separation in a manner reminiscent of the interactive 
behaviour of the pressure-gradient required in traditional triple-deck theory \cite[e.g.][]{Ruban}.

A number of mixed convection scenarios have been discussed in which a 
flow due to thermally-induced pressure gradient achieves a balance with a background flow field without
provoking flow separation. Typically these have been described using similarity structures of the stagnation-point
type which share many features with traditional boundary-layer theory but which do not formally require
approximation on the assumption of a small dimensionless parameter. Amin \& Riley \cite{AR95} found that
a steady solution may be constructed for mixed convection at a stagnation-point on a plane wall when 
the wall temperature gradient is assumed to vary quadratically with distance along the wall.
Surprisingly a steady solution is possible not only when thermally-driven flow component is in the same direction
as the background flow but also when it opposes the background flow.
Arrieta-Sanagust\'in \cite{AS11} considered the axisymmetric version of this problem
and demonstrated that a steady solution is possible with the thermal flow opposing the stagnation-point flow 
provided that the Grashof number lies below a threshold value. Above this value the assumed similarity structure fails and 
an eruption of fluid from the vorticity layer adjacent to the wall is anticipated. Further examples are
given by Revnic {\it et al.} \cite{RGMP}, who examined
mixed convection near a stagnation point on a vertical cylinder, and Merkin \& Pop \cite{MPOP}
who studied mixed convection at a vertical surface
exposed to a far-field shear flow. In both cases steady solutions are available when the thermal flow
reinforces the shear flow. Solutions also exist for an opposing thermal flow if a dimensionless mixed convection 
parameter is below a critical value. Finally we mention the very recent work of Riley \cite{Riley2013}, who 
examines the unsteady flow above a horizontal wall induced by a standing temperature wave.
 
In this paper we discuss mixed convection above a heated rotating disk. 
In the absence of a temperature field, the flow is described by a set of ordinary differential equations
according to the classical three-dimensional stagnation-point structure of von K\'arm\' an \cite[e.g.][]{Schi}.
Solutions of these equations have been studied exhaustively. Cochran \cite{Cochran1934} computed the most 
well-known solution, and that which is most commonly described in textbooks \cite[e.g.][]{Schi, Batchelor}. 
More recently, a number of authors have pointed out the existence of multiple solutions. Zandbergen and Dijkstra
\cite{Zandbergen1977} computed a second solution as part of a wider study which accounted for the effect 
of suction at the plate. Later, Lentini and Keller \cite{Lentini1980} showed that in fact there are an infinite number
of solutions. An excellent survey of the results is given by Zandbergen and Dijkstra \cite{Zandbergen1987}.
In the presence of a temperature field, the flow can be analysed using the same von K\'arm\'an similarity
structure with the additional assumption that the wall temperature varies quadratically
with distance away from the centre of the disk. 
The mixed convection is characterised by two dimensionless parameters, a Prandtl number and a Grashof number.
The rotation of the disk draws fluid downwards from infinity
and forces it outwards away from the disk centre. For positive Grashof number, 
the thermally-induced pressure gradient opposes this and
attempts to drive fluid towards the axis of rotation. For negative Grashof number, the thermally-induced
pressure gradient complements the centrifugal flow. This case was studied by Sreenivasan \cite{sree}.

Section 2 is devoted to steady solutions for a fixed Prandtl number, 
which are expected to exist when the Grashof number lies below a critical value. Moreover
the non-uniqueness mentioned above for isothermal flow strongly suggests the presence of multiple solution branches for 
the heated flow problem, and this is indeed found to be the case. 
The steady solutions are investigated further in the limit of small and large Prandtl number in 
section 3. Above the critical Grashof number the solution to the unsteady, self-similar form of the governing equations 
encounters a finite time singularity, and this is explored numerically in section 4. Our findings are summarised in section 5.



\section{Steady flow}\label{sec:steady}

We consider the steady flow generated by a combination of a horizontal, rotating flat plate and a thermally-induced
radial pressure gradient. The flow is assumed to be incompressible and satisfy the
Navier-Stokes equations under the Boussinesq approximation. The rotating plate boundary is
located at $z^*=0$ and the fluid occupies the region $z^*>0$. The plate is assumed to rotate about
the $z^*$ axis with angular velocity $\Omega\mathbf{k}$, where $\mathbf{k}$ is the unit vector in
the positive $z^*$ direction. 
The flow is assumed to be axisymmetric with velocity components $u^*,w^*$ in the $r^*,z^*$ directions
respectively  and $v^*$  is the azimuthal velocity component. The pressure in the fluid is
represented by $p^*$ and the temperature field in the fluid is represented by $\theta^*$. 
The plate temperature is assumed to vary quadratically with distance from the
origin so that 
\be
\theta^{*}=\theta_0^*-\left(\frac{b}{\nu/\Omega}\right )(\theta_0^*-\theta_\infty^*)\:r^2
\ee
at $z^*=0$, where $b$ is a positive dimensionless parameter, $\theta^*_\infty$ is the
ambient temperature far from the plate and $\theta^*_0$ is the plate temperature at the origin.

We non-dimensionalize by writing 
\begin{align}
&(r^*,z^*)=(\nu/\Omega)^{1/2}(r,z),\quad (u^*,v^*,w^*)=\left(\nu\Omega\right)^{1/2}(u,v,w),\\
&p^*=(\rho \nu\Omega) p,\quad
\theta^*= \theta^*_\infty+(\theta^*_0-\theta^*_\infty)\theta,\label{char_sc}
\end{align}
where 
$\nu$ is the kinematic viscosity. 
According to the Boussinesq approximation the fluid density is given by
\begin{equation}
\rho=\rho_\infty\Big(1-\beta(\theta^*_0-\theta^*_\infty)\theta\Big),
\end{equation}
where $\rho_\infty$ is the
density attained at the ambient temperature $\theta_{\infty}^*$ and $\beta$ is the coefficient of thermal
expansion far from the plate. 
The temperature field inside the fluid is governed by the dimensionless energy equation
\begin{equation}
\pr(u\theta_r+w\theta_z)=\theta_{rr}+\theta_r/r+\theta_{zz},
\label{energy_equation}
\end{equation}
where $\pr=\nu/\kappa$ is the Prandtl number and $\kappa$ is the thermal diffusivity of the fluid.

We seek a similarity solution of the form
\begin{equation}
u=r u_0(z),\quad v=r v_0(z),\quad w=w_0(z),\quad p=p_0(z)+r^{2}p_1(z),\quad
\theta=\theta_0(z)+br^{2}\theta_1(z).
\label{sim_sol}
\end{equation}
Substituting 
\eqref{sim_sol} into the Navier-Stokes equations we obtain  
\begin{align}
2 u_0+\frac{\dd w_{0}}{\dd z}&=0,\label{cont_eq_adim}\\
u_{0}^{2}-v_{0}^{2}+w_0\frac{\dd u_0}{\dd z}&=-2p_1+\frac{\dd^2 u_0}{\dd z^2},\label{rad_mom_eq_adim}\\
2 u_0 v_0+w_0\frac{\dd v_0}{\dd z}&=\frac{\dd^2 v_0}{\dd z^2},\label{tangential_mom_eq_adim}\\
2 u_0\theta_1+w_0\frac{\dd \theta_1}{\dd z} &=  \frac{1}{\sigma} \frac{\dd^2 \theta_1}{\dd z^2}.\label{energy_adim}
\end{align}
The boundary conditions are 
\begin{equation}
u_0=0,\;v_0=1,\;w_0=0, \quad \theta_1=-1 \quad \hbox{at} \quad z=0,
\label{bc_1}
\end{equation}
and
\begin{equation}
u_0, \,v_0,\,\theta_1\to 0 \quad \hbox{as} \quad z\to\infty.
\label{bc_2}
\end{equation}
The pressure field $p_1$ is given by 
\begin{equation}
p_1=-\gr\int_z^{\infty}\theta_1\,\mathrm{d}z,
\label{p_1_eq}
\end{equation}
where the dimensionless Grashof number is defined to be 
\begin{equation}
\gr=\frac{\beta g  (\theta_0^*-\theta_\infty^*)}{\nu^{1/2}\Omega^{3/2}}.
\end{equation}
The remaining unknown functions in (\ref{sim_sol}), namely $p_0(z)$ and $\theta_0(z)$, satisfy a pair of ordinary differential equations, with accompanying
boundary conditions, which can be written down in a straightforward manner. 
These are not the main focus of the present work, and so we do not give these here;
our primary concern is with the system
\eqref{cont_eq_adim}-\eqref{p_1_eq}. 


The problem is solved numerically by first truncating the infinite domain of integration at a finite level,
$z_\infty$, and introducing a non-uniform grid covering the range $[0,z_\infty]$. In pratice it is
useful to cluster grid points at $z=0$ in order to more accurately resolve the
solution in the region close to the wall where variation is typically rapid. The equations are discretized using centered 
differences for the derivative terms at each interior grid point. The boundary conditions \eqref{bc_1} are enforced
at the first grid point and the conditions in
\eqref{bc_2} together with $p_1=0$ are enforced at the last grid point. The
non-linear terms in the governing equations are dealt with using a quasi-linearization
procedure as part of an iterative process. For example the term $u_0^2$ in \eqref{rad_mom_eq_adim} is replaced by
$2u_0\tilde{u}_0-\tilde{u}_0^2$ where $\tilde{u}_0$ is the value from the previous iteration. In
this way at each iteration we solve a tri-diagonal system of linear algebraic equations for the
unknowns at each grid point. This is done efficiently using the Thomas algorithm \cite[e.g.][]{POZ}. The solution is deemed 
to have converged once the condition $|u_0-\tilde{u}_0|<\delta$ is met at every grid point
for a prescribed choice of $\delta$. A similar convergence condition is required for all of the other flow variables.
In practice we took $\delta=10^{-8}$. As a check, the converged numerical solution is
substituted back into the governing equations.

When $\gr=0$, so that thermal effects are absent, the problem reduces to the
classical von K\'arm\'an rotating disk flow for which fluid is swept radially outwards along the plate and drawn down
from infinity along the axis of rotation. For $\gr>0$ a thermally-induced pressure
gradient tends to force fluid radially inwards and up away from the plate.
The presently sought similarity solution
assumes that the magnitude of the vorticity decays with vertical distance upwards from the boundary. Its existence
will be determined by a competition between the centrifugally-driven downwash, which tends to draw vorticity towards the
boundary, and the thermally-driven upwash which tends to carry it away from the boundary. If $\gr$ is sufficiently large
we expect the latter to overcome the former and, consequently, for the hypothesised similarity
structure to fail.


\begin{figure}[!t]
\begin{center}
\includegraphics[width=0.7\textwidth]{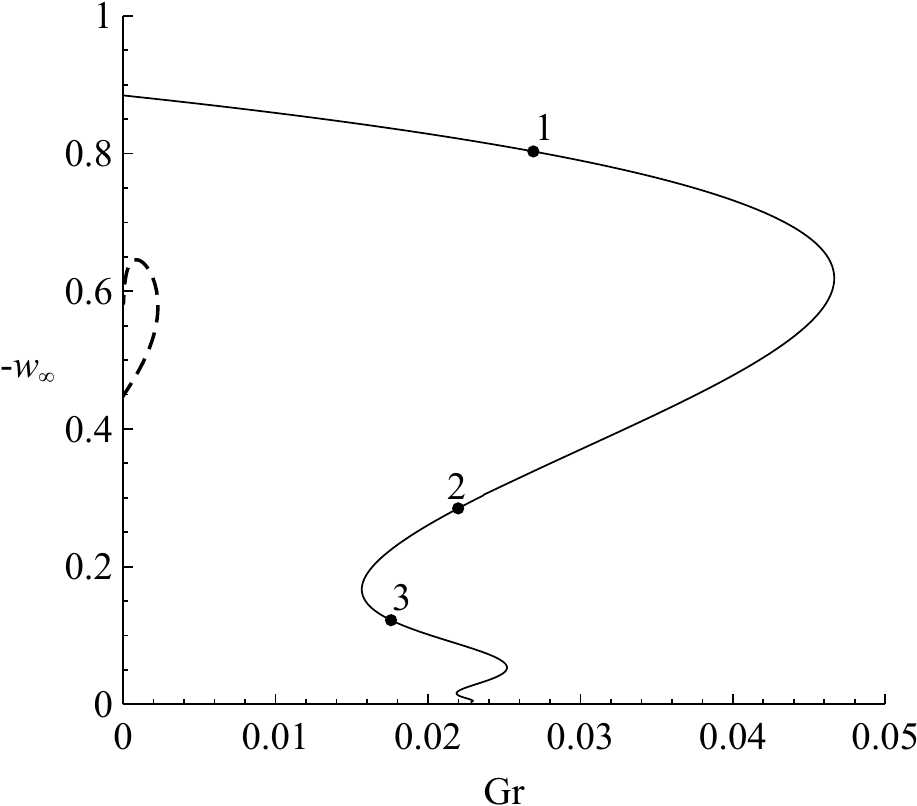}
\caption{The steady solution curves for $\pr=1$. The main branch is shown with a solid line, and one of the secondary branches is shown with a broken line. Velocity and temperature profiles at the points labelled 1, 2, 3 are shown in figures
\ref{prof1}, \ref{prof2} and \ref{prof3} respectively.}
\label{winf_vs_gr}
\end{center}
\end{figure}
As was mentioned in the Introduction, the presence of multiple solution branches is expected.
In Fig. \ref{winf_vs_gr} we show two of the solution branches for the case $\pr=1$. The flow is characterised
using the negative of the vertical velocity at infinity, $w_\infty$. On both branches the
solution at $\gr=0$ was computed using the numerical method described 
above. The branches were then extended to positive Grashof numbers using the parameter
continuation software AUTO-07p \cite{Doed} with the finite difference solution at 
$\gr=0$ used as a starting point. The branch shown with a solid line starts at the point 
$(\gr,w_\infty)=(0,0.884)$, which coincides with the appropriate calculation of Zandbergen 
and Dijkstra \cite{Zandbergen1977} (see their table 4.1). The branch shown with a broken
line starts at $(\gr,w_\infty)=(0,0.447723)$, which also agrees with table 4.1 of Zandbergen 
and Dijkstra \cite{Zandbergen1977}.
As the branch extends away from $\gr=0$ and starts to oscillate down toward the $\gr$ axis, we found that
it was necessary to gradually increase $z_\infty$ as the thickness of the vorticity layer at the disk grows. 
Over most of the branch we took $z_\infty=500$, but towards the lowermost part of the branch shown in the figure
it was necessary to take the very large value $z_\infty=2\times 10^4$.
For practical reasons, the computations were terminated at the end-point shown. It seems reasonable to suppose
that the branch will continue to approach the $\gr$ axis while the thickness of the vorticity layer continues to grow.

Having delineated the main branch as described, a few individual points along the branch were confirmed 
independently using an alternative numerical method as follows. 
The value of $w_{\infty}$ was fixed and $\gr$ was treated as an 
unknown to come as part of the solution.  Starting from the large $z$ asymptotic form 
of the solution provided by \cite{Zandbergen1977}, \eqref{cont_eq_adim}-\eqref{energy_adim} 
were integrated inwards from infinity, and a shooting method was 
used to satisfy the conditions at the disk \eqref{bc_1}. In each case the resulting pair
$(\gr,-w_{\infty})$ was confirmed to lie on the main branch in Fig. 
 \ref{winf_vs_gr}.

\begin{figure}[!t]
\begin{center}
\includegraphics[width=0.7\textwidth]{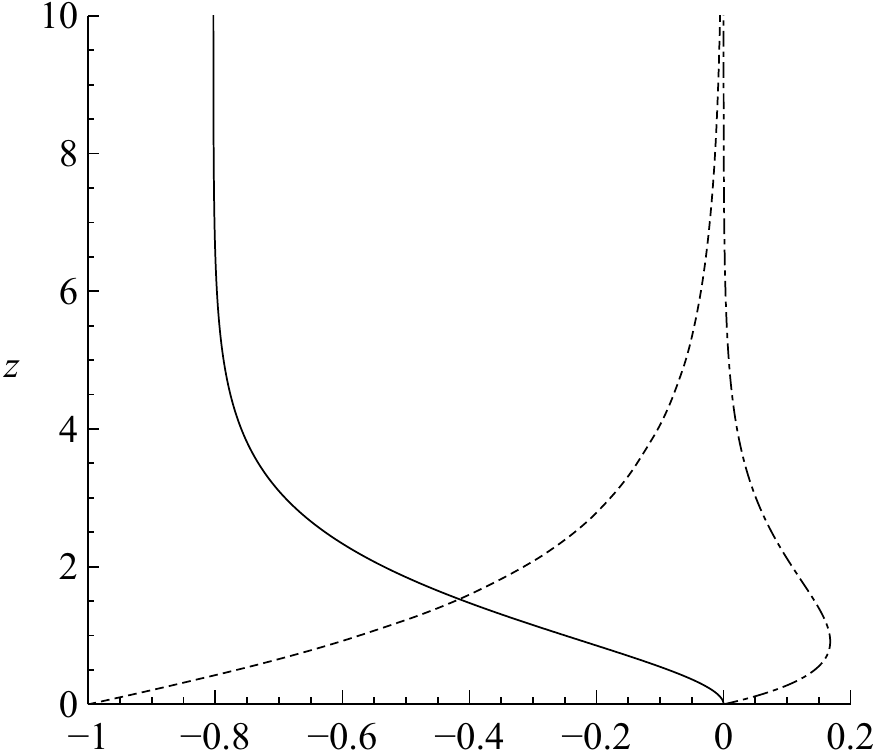}
\caption{The velocity profiles $w_0$ (solid line), $u_0$ (dot-dashed line), and the temperature profile $\theta_1$ (dashed line) for $\pr=1$ and $\gr=0.0269$, corresponding to label 1 in figure \ref{winf_vs_gr} with $w_\infty=-0.803$. Note that for $\pr=1$, as here, $v_0=-\theta_1$.}
\label{prof1}
\end{center}
\end{figure}

\begin{figure}[!t]
\begin{center}
\includegraphics[width=0.7\textwidth]{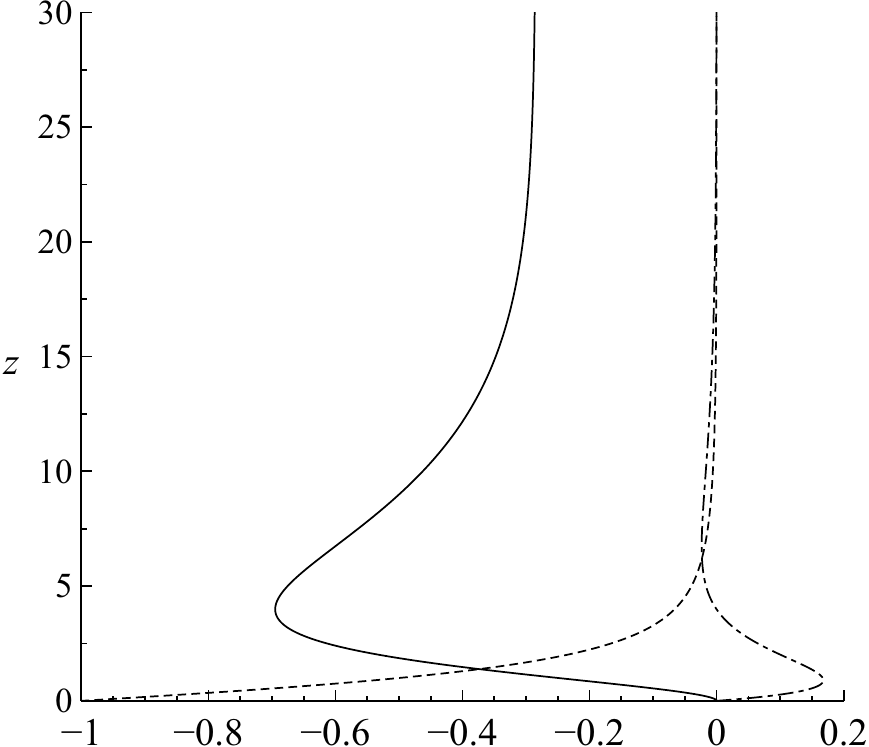}
\caption{The velocity profiles $w_0$ (solid line), $u_0$ (dot-dashed line), and the 
temperature profile $\theta_1$ (dashed line) for $\pr=1$ and $\gr=0.0219$, corresponding to label 2 in figure \ref{winf_vs_gr} with $w_\infty=-0.284$. Note that for $\pr=1$, as here, $v_0=-\theta_1$.}
\label{prof2}
\end{center}
\end{figure}

\begin{figure}[!t]
\begin{center}
\includegraphics[width=0.7\textwidth]{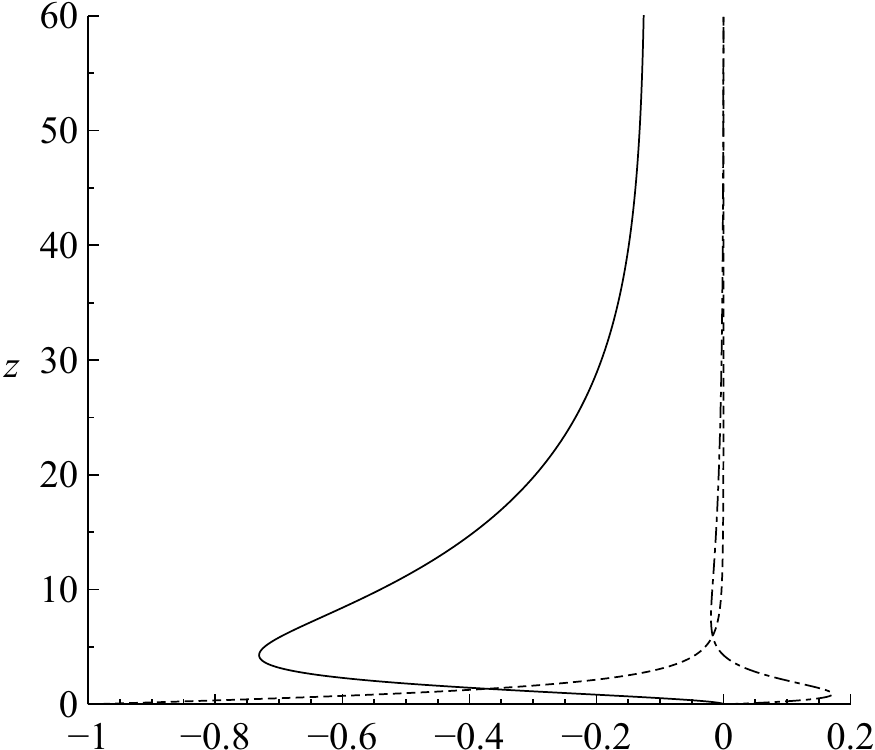}
\caption{The velocity profiles $w_0$ (solid line), $u_0$ (dot-dashed line), and the 
temperature profile $\theta_1$ (dashed line) for $\pr=1$ and $\gr=0.0176$, corresponding to label 3 in figure \ref{winf_vs_gr} with $w_\infty=-0.122$. Note that for $\pr=1$, as here, $v_0=-\theta_1$.}
\label{prof3}
\end{center}
\end{figure}

On the main solution branch shown in Fig. \ref{winf_vs_gr}, 
there is one solution when $0<\gr<0.0156$, there are three 
solutions when $\gr>0.0156$, and there are no solutions when $\gr>\gr_c$, where the critical
Grashof number $\gr_c=0.0466$. Velocity 
and temperature profiles at three locations on the main branch are shown in Figs
\ref{prof1}, \ref{prof2} and \ref{prof3}. Note that in the presently considered case of $\pr=1$, it is readily seen from the 
governing equations \eqref{tangential_mom_eq_adim} and \eqref{energy_adim}
and the boundary conditions \eqref{bc_1} and \eqref{bc_2} that $v_0=-\theta_1$
and so the profile for the swirl component of velocity may be easily inferred from the temperature profiles shown in the 
figures.
In each case the vertical velocity is single-signed and the fluid descends from infinity toward the disk. Notably, the radial 
velocity switches sign in the profiles shown in Figs \ref{prof2} and \ref{prof3} as a turning point develops in the axial 
$w_0$ profile.
In Figs \ref{prof1}, \ref{prof2}, and \ref{prof3}, $\theta_1$, and hence also 
the swirl component $v_0$, has the same sign everywhere.

Other values of the Prandtl number, $\pr$, were also investigated, and broadly the same behaviour was found as that 
discussed above. For example, the main solution branch was computed for the case $\pr=0.5$ and it is found to be have 
qualitatively the same shape as that shown in Fig. \ref{winf_vs_gr}.
Fig. 5 shows how the critical Grashof number beyond which there are no steady solutions
depends on the Prandtl number, $\sigma$.
Notably, as $\sigma$ increases, $\gr_c$ appears to grow without bound, and as $\sigma$ decreases, 
$\gr_c$ appears to approach zero. These trends are investigated further in the next section where we examine
the limits of large and small $\sigma$ in turn.



\begin{figure}[!t]
\begin{center}
\includegraphics[width=0.8\textwidth]{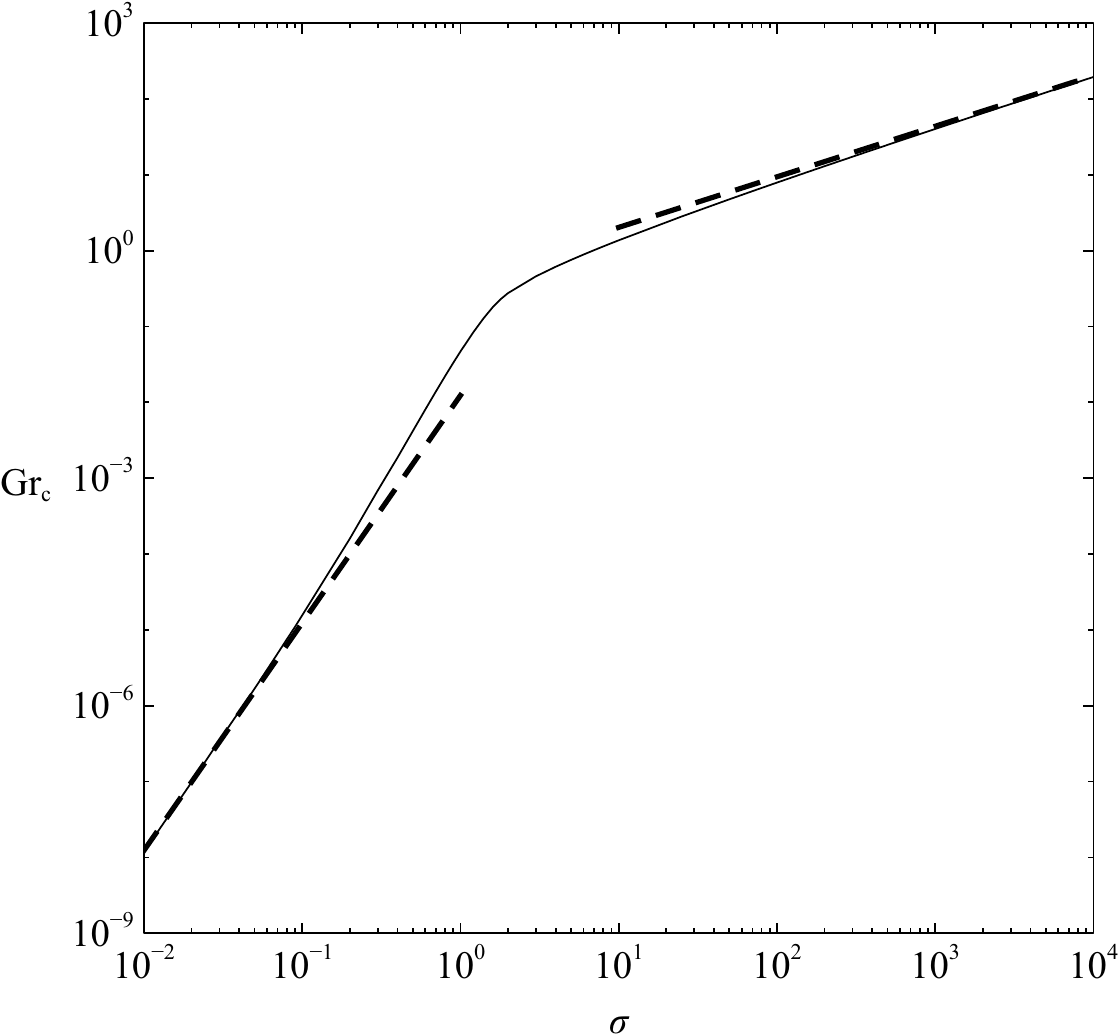}
\caption{The dependence of the critical Grashof number, $\gr_c$, on the Prandtl number $\sigma$, shown with 
a solid line. The broken lines correspond to the asymptotic predictions,
$\gr_c=0.436\sigma^{2/3}$ in (\ref{papoose}) and 
$\gr_c=0.0123\sigma^{3}$ in (\ref{grasp}), 
for large and small $\sigma$ respectively.}
\label{gr_c_large_sigma}
\end{center}
\end{figure}



\section{Steady flow for large and small Prandtl number}

Here we use asymptotic analysis to discuss the flow when $\sigma \gg 1$ and $\sigma\ll 1$. Of particular
interest is the behaviour of the the critical Grashof number, $\gr_c$, in these limits.

\subsection{The case $\sigma\gg1$}

For large Prandtl number the diffusivity of heat from the boundary is much less than the diffusivity of vorticity created at the 
boundary. A consequence of this is that the thermal boundary layer is much thinner than the velocity boundary layer and 
so within the inner, thermal boundary layer the radial velocity is a linear function of $z$, i.e. $u_0\propto z$. A 
consequence is that if convection and diffusion of heat are to balance 
therein, \eqref{energy_adim} shows that the thickness of this inner layer is 
$O(\sigma^{-{1\over3}})$, a classic result. Further $u_0=O(\sigma^{-{1\over3}})$, and 
the continuity equation \eqref{cont_eq_adim} requires $w_0=O(\sigma^{-{2\over3}})$, with $v_0\equiv1$ at leading order. A balance 
between the pressure and viscous forces in the radial momentum equation \eqref{rad_mom_eq_adim} 
then requires $p_1=O(\sigma^{1\over3})$. Finally we note from equation \eqref{p_1_eq} that Grashof numbers as large 
as $O(\sigma^{2\over3})$ are available for steady flow in this situation. For this inner region then we introduce the scaled variables
\begin{equation}\label{pooch}
u_0=\sigma^{-{1\over3}}U_0, \qquad w_0=\sigma^{-{2\over3}}W_0, \qquad p_1=\sigma^{1\over3}P_1, \qquad Gr=\sigma^{2\over3}\Lambda, \qquad 
z=\sigma^{-{1\over3}}Z,
\end{equation}
which yield the following equations for the inner layer
\begin{equation}
2U_0 + \frac{\dd W}{\dd Z}=0, \qquad 0=-2P_1 + \frac{\dd^2 U_0}{\dd Z^2}, \label{herring}
\end{equation}
\begin{equation}
2U_0\theta_1+W_0\frac{\dd \theta_1}{\dd Z} = \frac{\dd^2\theta_1}{\dd Z^2}, \qquad \frac{\dd P_1}{\dd Z}=\Lambda\theta_1,
\end{equation}
with boundary conditions
\begin{equation}\label{desk}
U_0=W_0=0, \quad \theta_1=-1 \quad \hbox{at} \quad Z=0, \quad P_1,\theta_1\to 0\quad \hbox{as} \quad Z\to\infty,
\end{equation}
together with a suitable matching condition for $U_0$ as $Z\to\infty$. 

Outside this thin thermal boundary layer $p_1=\theta_1=0$ and the velocity components are each of order unity. 
Therefore, the governing equations for the outer layer, whose characteristic thickness is of order unity, are from (\ref{cont_eq_adim}) to (\ref{tangential_mom_eq_adim})  
\begin{equation}
2u_0 + \frac{\dd w_0}{\dd z} = 0, \quad u_0^2 - v_0^2 + w_0 \frac{\dd u_0}{\dd z} = \frac{\dd^2 u_0}{\dd z^2}, \quad 2u_0v_0+w_0\frac{\dd v_0}{\dd z} = \frac{\dd^2 v_0}{\dd z^2},
\end{equation}
together with 
\begin{equation}
u_0=v_0-1=w_0=0 \quad \hbox{at} \quad z=0, \quad u_0,v_0\to 0 \quad \hbox{as} \quad z\to\infty.
\end{equation}
Therefore the flow in this outer layer corresponds to the classical von K\'arm\' an rotating disk flow. In order to match the inner and outer 
solutions note from (\ref{herring}) 
that as $Z\to\infty$, $U_0\propto Z$. Accordingly, taking account of the scalings in (\ref{pooch}), the matching 
requirement is 
\begin{equation}
\left. \frac{\dd U_0}{\dd Z}\right|_{Z\to\infty} = \left.\frac{\dd u_0}{\dd z}\right|_{z=0}.
\end{equation}
The system of equations (\ref{herring}) to (\ref{desk}) has been integrated numerically using the method outlined in section 2 for varying 
values of $\Lambda$. In particular we find that no solutions exist for $\Lambda=\Lambda_c>0.436$. Therefore the asymptotic 
behaviour of the critical Grashof number for large values of the Prandtl number has
\begin{equation}\label{papoose}
\gr_c\sim0.436\sigma^{2\over3} \quad \hbox{as} \quad \sigma\to\infty.
\end{equation}

{\no}This result is compared with the results obtained from the numerical solutions of (\ref{cont_eq_adim}) to (\ref{p_1_eq}) in Fig. 5, showing good agreement for large 
values of $\sigma$.\medskip

\subsection{The case $\sigma\ll1$}

For very small values of the Prandtl number, the thermal diffusivity is much greater than the 
diffusivity of vorticity, with the consequence that the thermal boundary layer now 
exceeds, by far, the velocity boundary layer in thickness.
Assuming, as is reasonable, that the range of available values of $\gr$ is small (see Fig. 5)
in the inner region, where $z=O(1)$, we have from (\ref{cont_eq_adim})-(\ref{energy_adim}), at leading order
\begin{equation}\label{anne}
2\tilde{u}_0+\frac{\dd \tilde{w}_0}{\dd z}=0, \quad \tilde{u}^2_0-\tilde{v}^2_0+\tilde{w}_0{\dd\tilde{u}_0\over \dd z}={\dd^2\tilde{u}_0\over \dd z^2}, \quad 
2\tilde{u}_0\tilde{v}_0+\tilde{w}_0{\dd\tilde{v}_0\over \dd z}={\dd^2\tilde{v}_0\over \dd z^2},  
\end{equation}
together with
\begin{equation}\label{boleyn}
\tilde{u}_0=\tilde{v}_0-1=\tilde{w}_0=0 \quad \hbox{at} \quad z=0, \quad \tilde{u}_0,\tilde{v}_0\to 0
\end{equation}
as $z \to \infty$. 
Notice that we have used a tilde superscript to indicate variables in the inner region.
System (\ref{anne}), (\ref{boleyn}) is uncoupled from the temperature field and is
identical to that which governs classical von K\'arm\' an flow over a rotating disk. The 
solution has the property that $\tilde{w_0}\to-k$ as $z\to\infty$, where $k=0.884$ (e.g. \cite{Zandbergen1987}, table 4.1), 
and $\tilde u_0$, $\tilde v_0\to 0$ exponentially fast. 

In the outer thermal boundary layer, we expect a balance between the outward diffusion and the inward convection of heat
so that in (\ref{energy_adim})
\begin{equation} \label{carrot}
{\dd^2\theta_1\over \dd z^2} \sim \sigma w_0 \frac{\dd \theta_1}{\dd z} = O\left (\sigma k{\dd \theta_1\over \dd z}\right).
\end{equation}
This suggests that the aforementioned balance occurs in a region of scale $O(\sigma^{-1})$ which prompts 
the definition of an outer variable $\zeta=\sigma z$. 
Written in terms of the new outer variable, 
(\ref{rad_mom_eq_adim}) and (\ref{p_1_eq}) become respectively,
\begin{equation}\label{cosine}
u^2_0-v^2_0-\sigma k{\dd u_0\over \dd\zeta}=-2p_1+\sigma^2{\dd^2u_0\over \dd\zeta^2}, \qquad
\frac{\dd p_1}{\dd \zeta} = \frac{\gr}{\sigma}\, \theta_1.
\end{equation}
The first of these suggests viscous diffusion is unimportant and, balancing the three 
momentum terms on the left hand side with the pressure term on the right hand side, indicates that 
$u_0=O(\sigma)$, $v_0=O(\sigma)$ and $p_1=O(\sigma^2)$ in this outer region. Substituting the last of these
scalings 
into the second equation in (\ref{cosine}), and assuming that $\theta_1=O(1)$, we deduce that
$\gr=O(\sigma^3)$. Accordingly we now write $\gr=\lambda\sigma^3$ where $\lambda=O(1)$ and seek to 
determine the critical value of $\lambda$ beyond which there is no steady solution.

With the above preliminaries in place we now develop the solution in greater detail. In the inner region we expand the 
velocity, pressure and temperature fields by writing
\begin{align}
\nonumber
&u_0=\tilde{u}_0+\sigma\tilde{u}_1+\sigma^2\tilde{u}_2+\cdots, \qquad v_0=\tilde{v}_0+\sigma\tilde{v}_1+\sigma^2\tilde{v}_2+\cdots,\\
\label{foot}
& w_0=\tilde{w}_0+\sigma\tilde{w}_1+\sigma^2\tilde{w}_2+\cdots, \qquad p_1 = \sigma^2 \tilde p_0 + \cdots, \qquad
\theta_1 = \tilde \theta_1 + \sigma \tilde \theta_2 +  \cdots.
\end{align}
Introducing these expansions into (\ref{cont_eq_adim})-(\ref{energy_adim}) we obtain (\ref{anne}) at leading order for which, as was mentioned above, the classical von K\'arm\' an solution is available. 
The leading order temperature field satisfies
\bea
\frac{\dd^2 \tilde\theta_1}{\dd z^2} = 0,
\eea
with $\tilde\theta_1=-1$ on $z=0$. 
Integrating and applying the wall boundary condition we find
$\tilde \theta_1 = -1 + Az$ for constant $A$. We find that a self-consistent solution can be constructed
by taking $A=0$ so that $\tilde\theta_1=-1$.
At first order, namely $O(\sigma)$, we obtain from (\ref{cont_eq_adim}) and (\ref{rad_mom_eq_adim}) respectively,
\begin{equation}\label{monkey}
2\tilde{u}_1+{\dd \tilde{w}_1\over \dd z}=0, \quad 2\tilde{u}_0\tilde{u}_1-2\tilde{v}_0\tilde{v}_1+\tilde{w}_0{\dd \tilde{u}_1\over 
\dd z}+\tilde{w}_1{\dd \tilde{u}_0\over \dd z}={\dd^2\tilde{u}_1\over \dd z^2},
\end{equation}
and from (\ref{tangential_mom_eq_adim}) we have 
\begin{equation}\label{mice}
2\tilde{u}_0\tilde{v}_1+2\tilde{u}_1\tilde{v}_0+\tilde{w}_0{\dd\tilde{v}_1\over \dd z}+\tilde{w}_1{\dd\tilde{v}_0\over \dd z}={\dd^2\tilde{v}_1\over \dd z^2}.
\end{equation}
At the wall we have $\tilde u_1=\tilde v_1=\tilde w_1=0$.
Taking the limit $z\to\infty$, and considering the dominant balances of terms in (\ref{monkey}), (\ref{mice}) we find
\begin{align}\label{cows}
\tilde{u}_1\sim \alpha_{1} + \edt \qquad
\tilde{w}_1\sim -2\alpha_{1}z + \alpha_{2} + \edt, 
\qquad \tilde{v}_1\sim \alpha_{3} + \edt
\end{align}
as $z\to \infty$, where $\edt$ means `exponentially decaying terms' and 
$\alpha_{1}$, $\alpha_{2}$, and $\alpha_{3}$ are constants.

\begin{figure}[!t]
\begin{center}
\includegraphics[width=0.7\textwidth]{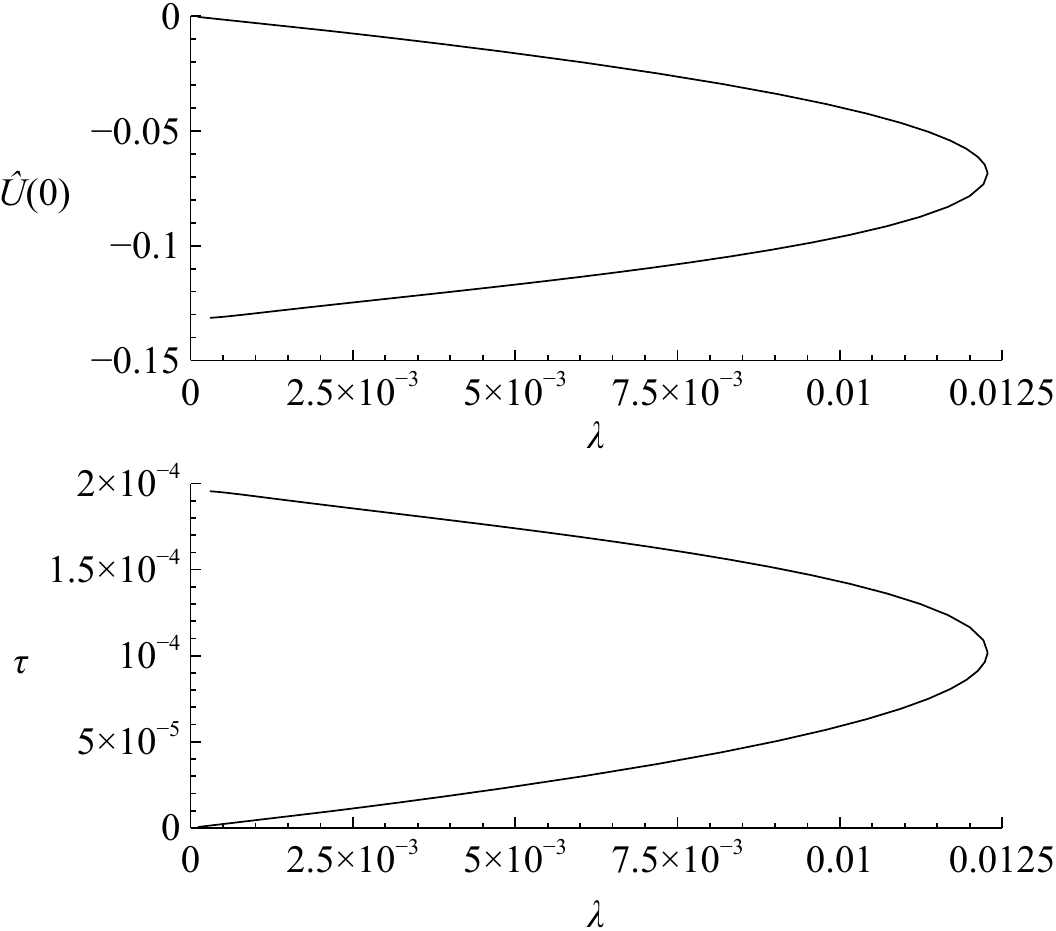}
\caption{$\sigma\ll 1$: The solution branches for the first order problem in the inner region 
(lower panel, showing $\tau \equiv \dd \tilde u_1/\dd z|_{z=0}$)
and the leading order problem in the outer region (upper panel, showing $\hat U(0)$).} 
\end{center}
\label{fig:asym}
\end{figure}
In the outer region, we expand by writing
\begin{align}
& u_0=\sigma \:\hat U(\zeta) + \cdots, \qquad v_0 = \sigma\: \hat V(\zeta) + \cdots, \qquad w_0 = \hat W(\zeta) + \cdots, \\
& p_1 = \sigma^2\:\hat P(\zeta) +\cdots, \qquad \theta_1 = \hat \Theta(\zeta) + \cdots,
\end{align}
where $\zeta = \sigma z$.
Substituting into (\ref{cont_eq_adim})-(\ref{energy_adim}) and (\ref{p_1_eq}) we obtain at leading order,
\begin{align}
\label{house}
2\hat{U}+{\dd\hat{W}\over \dd\zeta}=0, \qquad &\hat{U}^2-\hat{V}^2+\hat{W}{\dd\hat{U}\over \dd\zeta}=-2\hat{P}, 
\\
\label{looper}
2\hat{U}\hat{V}+\hat{W}{\dd\hat{V}\over \dd\zeta}=0,
\qquad 
&{\dd\hat{P}\over \dd\zeta} = \lambda\hat{\Theta}, \qquad 2\hat{U}\hat{\Theta}+\hat{W}{\dd\hat{\Theta}\over \dd\zeta}={\dd^2\hat{\Theta}\over 
\dd\zeta^2},
\end{align}
together with
\begin{equation}\label{robin}
\hat{W}(0)=-k, \quad \hat{\Theta}(0)=-1, \quad \hbox{and} \quad \hat{U},\hat{V},\hat{P},\hat{\Theta}\to 0 \quad \hbox{as} \quad 
\zeta\to\infty.
\end{equation}
The first two conditions in (\ref{robin}) are required to match to the inner region. Moreover, matching
is completed by taking
\bea \label{match}
\alpha_{1}=\hat{U}(0), \qquad \alpha_{3}=\hat{V}(0).
\eea

Integrating the first equation in (\ref{looper}), with the use of the first equation in (\ref{house}),  
we find that $\hat V = -(\alpha_3/k)\hat W$, where the constant of
integration has been chosen to fulfil the matching condition in (\ref{match}). Substituting this result, 
the system (\ref{house})-(\ref{robin}) is solved numerically using a shooting method based on
fourth-order Runge-Kutta integration. The numerical solution shows that $\alpha_3=0$ so that $\hat V\equiv 0$.
%
%
The numerical solution is illustrated in the upper panel of Fig. 6
, where $\hat U(0)$ is plotted
against $\lambda$. Evidently two possible solutions exist over the range of $\lambda$ values shown, and there
is no solution beyond the critical value $\lambda_c=0.0123$. The lower panel of the same figure
shows the corresponding solution branch, showing $\dd \tilde u_1/\dd z|_{z=0}$ versus $\lambda$,
 for the first order inner system (\ref{monkey})-(\ref{cows}). It should be noted that while the solution
 for the swirl component $\hat V$ is identically zero in the upper region, the inner solution $\tilde v_1$
 varies smoothly from zero at $z=0$, through non-zero values, toward zero as $z\to \infty$.
According to the preceding remarks, 
we now have the asymptotic estimate for the 
critical Grashof number,
\begin{equation}\label{grasp}
\gr_c\sim0.0123\sigma^3 \quad \hbox{as} \quad \sigma\to 0.
\end{equation}
This is compared with the critical value of the Grashof number obtained from the full equations in Fig. 6, showing 
excellent agreement. 

In concluding this section it is worth reflecting on the rather unexpected flow structure which has been uncovered.
First, we highlight the somewhat surprising form of the inner layer expansions (\ref{foot}). As was noted
above, the leading order velocity field, namely $(\tilde u_0,\:\tilde v_0,\:\tilde w_0)$, coincides with that for classical von 
K\'arm\' an flow on a rotating disk. Deviation from this basic profile occurs as a result of 
a thermally-induced pressure gradient which arises from the spatially-dependent wall temperature profile.
It is notable that although this thermally-induced pressure gradient appears at $O(\sigma^2)$
in the inner region expansions (\ref{foot}), it is necessary to include velocity perturbations of size $O(\sigma)$.
These velocity perturbations are driven indirectly by thermal effects via
the radial slip velocity which is imposed upon the inner layer by the solution in the outer region, where the effect
of a thermally-induced pressure-gradient is felt at leading order. Finally we note that the present flow structure
is similar in vein to that uncovered by Riley \cite{Riley2013} in his study of the flow response of a semi-infinite fluid
above a horizontal wall with a standing wave temperature profile.


\section{Time-dependent calculations}

Following the remarks of the previous section, it appears that no steady solution exists when the Grashof number lies above a critical value, $\gr_c$, which depends on the size of the Prandtl number.
To explore the flow behaviour above the critical Grashof number, we consider the unsteady form of the governing equations \eqref{cont_eq_adim}-\eqref{p_1_eq}
with the time-dependent terms included,
\begin{align}
2 u_0+\frac{\pl w_{0}}{\pl z}&=0,\label{cont_eq_unst}\\
\frac{\pl u_{0}}{\pl t}+u_{0}^{2}-v_{0}^{2}+w_0\frac{\pl u_0}{\pl z}&=-2p_1+\frac{\pl^2 u_0}{\pl z^2},\label{rad_mom_eq_unst}\\
\frac{\pl v_{0}}{\pl t}+2 u_0 v_0+w_0\frac{\pl v_0}{\pl z}&=\frac{\pl^2 v_0}{\pl z^2},\label{tangential_mom_eq_unst}\\
\frac{\pl \theta_{1}}{\pl t} + 2 u_0\theta_1 + w_0\frac{\pl \theta_1}{\pl z}&=\frac{1}{\pr}\frac{\pl^2 \theta_1}{\pl z^2},\label{energy_unst}
\end{align}
where $t=\Omega t^*$ is a dimensionless time, together with \eqref{p_1_eq}. These are to be integrated forwards in time subject to the boundary conditions \eqref{bc_1} and \eqref{bc_2}. A suitable initial condition is given by
\begin{equation}
u_0=-\frac{1}{2}\frac{\dd F}{\dd z}, \qquad v_0=G, \qquad w_0=F, \qquad \theta_1=p_1=0 
\label{initial_cond}
\end{equation}
at $t=0$, where the functions $F(z)$ and $G(z)$ are Cochran's \cite{Cochran1934} solution to 
\eqref{cont_eq_adim}-\eqref{tangential_mom_eq_adim} at $\gr=0$. We note that the same
initial condition was used by Riley \cite{Riley1964,Riley1969b} and Andrews and Riley \cite{Riley1969a} 
in their studies of heat transfer on a rotating disk. 

\begin{figure}[!t]
\begin{center}
\includegraphics[width=0.85\textwidth]{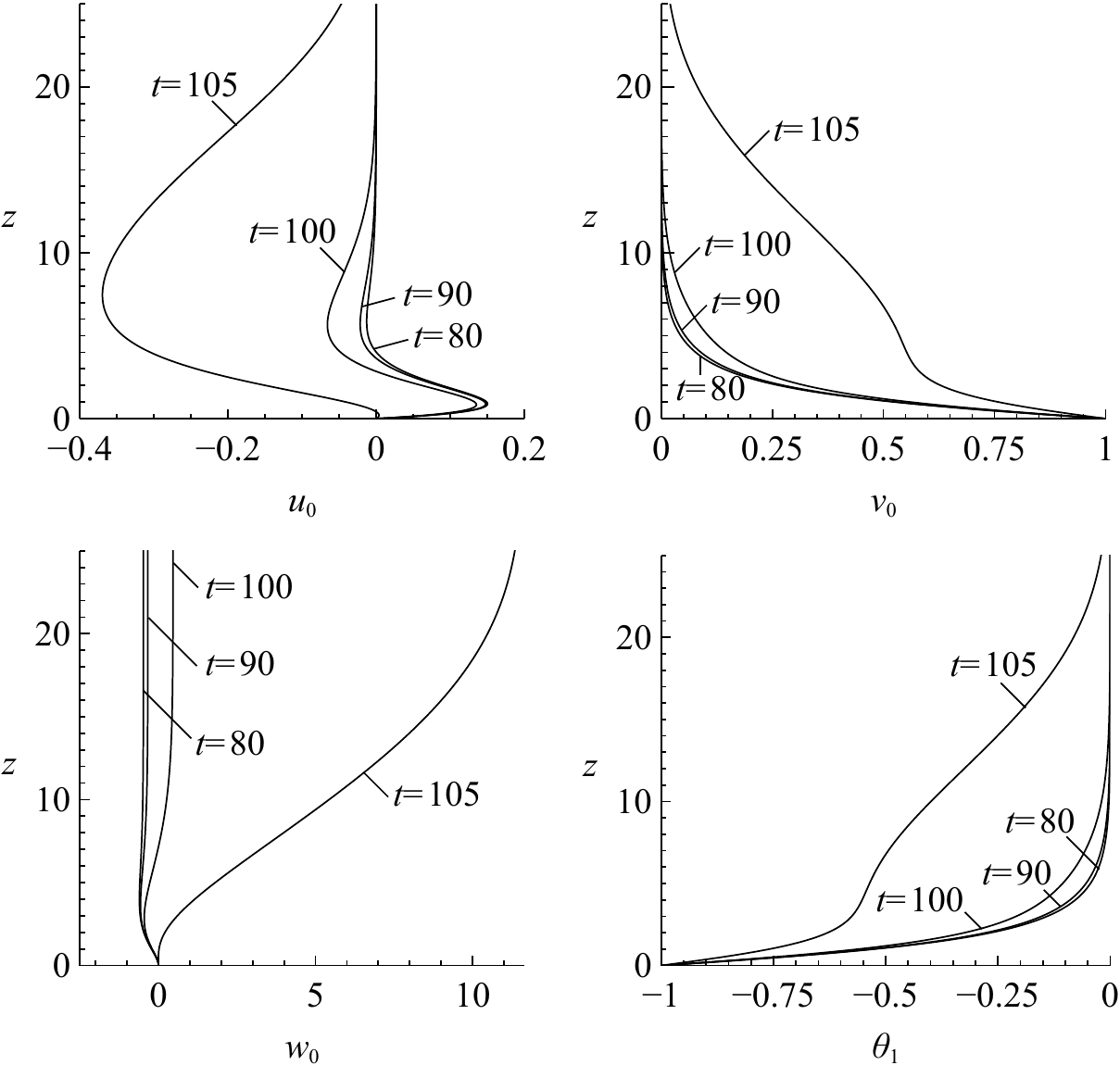}
\caption{Velocity and temperature profiles $u_0$, $v_0$, $w_0$, and $\theta_1$ at various dimensionless times for $\pr=1$ and $\gr=0.05$.}
\label{unsteady_gr_5d_2}
\end{center}
\end{figure}
The integration forwards in time may be achieved numerically employing a Crank-Nicolson scheme
based on the same discretisation of the equations as in Bodonyi and Stewartson \cite{Bodonyi1977}.
The nonlinear terms are dealt with at each time step 
using an iterative method similar to that described by Tanehill {\it et al.} \cite{Tanehill1984}.
A uniform time-step of size $0.01$ was used for the Crank-Nicolson scheme, whereas the same
spatial discretization as described in section \ref{sec:steady} was employed.

As in the previous section we focus attention on the case $\pr=1$. The velocity and temperature profiles $u_0$, $v_0$, $w_0$ and $\theta_1$ are shown in Fig. \ref{unsteady_gr_5d_2} at different times $t$. 
Similar behaviour is encountered at other Grashof numbers above the critical value.
In the early part of the flow evolution, the majority of the vorticity
is confined to a layer adjacent to the disk. As time increases, the thickness of this layer
increases as vorticity spreads further and further from the disk; this is evident from the behaviour of the $v_0$ curves in Fig. \ref{unsteady_gr_5d_2}, which approach zero further and further from the disk as time goes on.
Eventually, the numerical evidence is that a finite-time singularity is encountered 
and the solution blows up (in this case at $t\approx 107$). 
This singular behaviour lends strong support to the assertion that no steady solution exists 
above the critical value of the Grashof number discussed in section 2.





\section{Summary}

We have studied mixed convection over a rotating disk with the flow driven in part by the rotation of the disk
and in part by a thermally-induced pressure gradient which arises in response to a prescribed temperature
profile in the disk. Working on the basis of the Boussinesq approximation, we derived a set of partial differential
equations describing the flow and the temperature profile throughout the fluid according to an assumed similarity structure.
Steady solutions of this system are possible when the dimensionless Grashof number lies below a critical value which
depends on the size of the Prandtl number. Below the critical Grashof number, an infinite number of solution
branches exist, each one stemming from one of the infinity of branches which exist at zero Grashof number for 
isothermal flow. The main solution branch, which stems from the well-known Cochran solution of the classical
von K\'arm\'an problem at zero Grashof number, undergoes a number of successive turns so that 
multiple steady solutions exist along this branch alone. Asymptotic descriptions
of the flow, and in particular estimates for the critical Grashof number, were obtained 
in the limit of large and small Prandtl numbers and shown to be in agreement with the numerical
computations. Beyond the critical Grashof number there are no steady solutions; rather, our unsteady calculations
suggest that the solution of the unsteady similarity equations, subject to a suitable initial condition, terminates
in a finite-time singularity.

\section*{Acknowledgement}

The authors would like to thank Professor N. Riley for many helpful discussions during the preparation of this work.


 \end{document}